**An integrated and flexible ultrasonic device for the continuous bladder volume monitoring**

Alp Timucin Toymus[1], Umut Can Yener[1], Emine Bardakcı[2], Özgür Deniz Temel[2], Ersin Koseoglu[3], Dincay Akcoren[4], Burak Eminoglu[5], Mohsin Ali[6], Tufan Tarcan[3], Levent Beker*[1,6]

[1] Department of Mechanical Engineering, Koç University, Rumelifeneri Yolu, Sarıyer, Istanbul, 34450, Turkey

[2] Department of Electrical and Electronics Engineering, Koç University, Rumelifeneri Yolu, Sarıyer, Istanbul, 34450, Turkey

[3] Department of Urology, School of Medicine, Koç University, Rumelifeneri Yolu, Sarıyer, Istanbul, 34450, Turkey

[4] Analog Devices, Istanbul, Turkey

[5] Department of Electrical Engineering and Computer Sciences, University of California, Berkeley, Berkeley, CA, USA

[6] Department of Biomedical Sciences and Engineering, Koç University, Rumelifeneri Yolu, Sarıyer, Istanbul, 34450, Turkey

**Abstract**

Bladder volume measurement is critical for early detection and management of lower urinary tract dysfunctions. The current gold standard is invasive, and alternative technologies either require trained personnel or do not offer medical grade information. Here, we report an integrated wearable ultrasonic bladder volume monitoring (UBVM) device for accurate and autonomous continuous monitoring of the bladder volume. The device incorporates flexible and air-backed ultrasonic transducers and miniaturized control electronics with wireless data transmission capability. We demonstrated the real-life application of the device on healthy volunteers with various bladder shapes and sizes with high accuracy. Apart from the lower urinary tract dysfunctions, the proposed technology could also be adapted for various wearable ultrasonic applications.

**Introduction**

Lower urinary tract dysfunction (LUTD) is an umbrella term that includes conditions and diseases related to the bladder and the urethra[1]. LUTD affected 2.3 billion people worldwide in 2018, which is expected to increase in future due to the aging population[2]. Common conditions that cause LUTD include benign prostate hyperplasia (BPH)[3] and neurological disorders like multiple sclerosis and spinal cord injury[4,5]. For BPH-LUTD alone, the global economic burden was estimated to be more than $73 billion in 2019[6].

Since LUTDs revolve around the abnormalities during the filling and emptying of the bladder, bladder volume monitoring is an essential tool to evaluate bladder function[7]. At present, the gold standard measurement method is the bladder catheterization (Supplementary Note 1)[8]. Despite its accuracy, it is not recommended for routine examinations due to the high adverse outcomes associated the invasive procedure[9]. Hence, clinicians rely on ultrasound imaging of the bladder to estimate the bladder volume, which is a quick and non-invasive method to obtain information such as post-void residual volume of the bladder (Supplementary Note 2). However, this method cannot provide continuous monitoring and necessitates trained personnel with expensive benchtop equipment. Moreover, both of these methods are applicable to patients while they are admitted to the hospital only. Outside the hospital, monitoring relies on the patients themselves to fill urinary diaries and questionnaires to provide information regarding the volume and frequency of micturition[10]. Although continuous monitoring via implantable sensors does exist at the research level[11–15], only patients with complicated or severe symptoms would be suitable for such invasive devices[16]. Given the importance of bladder volume monitoring and the current status; precise and non-invasive technologies to monitor bladder volume outside the hospital settings are needed[17].

To this end, decades of research have led to several commercial wearable devices in which signals from the anterior and posterior walls of the bladder are used to quantify the degree of bladder filling[18–20]. Although such devices might be useful to alert the users when the bladder is full, they are not intended for accurate and continuous bladder volume measurement which is critical for diagnostic and treatment applications[21–23]. In addition, these rigid and non-conformal devices suffer from reliability issues and require ultrasound gel for proper signal transmission which limits their usage during daily activities[24,25]. Furthermore, although recent scientific studies reported improved accuracies for continuous volume measurement, these studies either incorporate bulky electronics subsystems or lack such systems totally[26–30]. Other reported methods include electrical impedance analysis[31–34] and near-infrared spectroscopy[35–37]. However, these methods could only provide preventative insights, and such indirect measurements would only be suitable for coarse estimations of the bladder volume due to their low specificity, proneness to motion artifacts, and urine dependent properties[38,39]. Therefore, none of these devices are suitable for accurate and continuous bladder volume monitoring.

A significant amount of research has been conducted in recent years on wearable ultrasonic transducers which has opened up a new dimension in the healthcare continuum[40–48]. Nevertheless, majority of the reported devices consist of ultrasonic transducers which require benchtop electronics and prevent monitoring outside hospital settings. Here, we introduce a wireless and flexible UBVM device with integrated electronics for continuous monitoring with high accuracy. By utilizing multiple transducers for A-mode ultrasonic measurements coupled with a low-power receiver circuitry and a spherical fitting algorithm for bladder volume estimation, the UBVM device enables wireless, non-invasive, and accurate measurement of the bladder volume continuously. Air-backed design of the ultrasonic transducers allows for accurate, gel-free and low-voltage bladder volume measurements of people of all ages. At the same time, a simple and effective time-of-flight measurement algorithm avoids the complex computations commonly associated with ultrasonic imaging and allows the integration of miniaturized electronics to realize a wearable ultrasonic monitoring system. Using a Bluetooth connection from the device to a mobile phone, bladder volume measurements can be monitored continuously. Wireless operation of the system is demonstrated through *in vitro* and *in vivo* experiments.

## Results and discussion

**Design and working principle of the integrated ultrasonic device.** The design concept and the operating principle of the UBVM device are illustrated in Fig. 1a. Once placed in the lower abdomen, the UBVM device performs continuous and wireless monitoring of the bladder volume. This is achieved by continuously recording the distance between the anterior and the posterior wall of the bladder from multiple sites using multiple transducers. These measurements are then transferred to a nearby mobile device and utilized in a volume estimation algorithm. The measurement process involves sequentially exciting the transducers, which emit ultrasound waves towards the bladder.

In our specific application, these ultrasound waves are reflected from the anterior and the posterior wall of the bladder. Upon returning to the transducers, the waves are converted back into electrical signals, carrying location information of the reflection points from the bladder wall. This information is then transferred to a mobile device via Bluetooth technology. With sequential repetition of such a measurement from multiple transducers at multiple sites, our algorithm estimates a sphere with its surface points corresponding to the reflection points on the bladder surface. The volume of the estimated sphere is then displayed on the custom mobile app we have developed, along with the measurement history. Further details regarding the operation of the device can be found in Methods and Supplementary Note 3.

The three main subsystems of the UBVM device are (1) a millimeter-scale conformal ultrasonic transducer module, (2) miniaturized electronics with integrated Bluetooth Low Energy capability, and (3) a custom mobile app that wirelessly communicates with the UBVM hardware and display the bladder volume measurements (Fig. 1a). In addition to PZT discs and their matching layers, the transducer module consists of a flexible substrate and soft silicone-based encapsulation layer to conform to the human skin (Fig. 1b,c). The module has a footprint of 3 x 3 $cm^2$. Moreover, to enable gel-free measurements, a silicone-based couplant was used to eliminate the air in between the transducers and the skin. From the transducers´ point of view, the UBVM device dramatically reduces the required voltage levels by utilizing an air-backed design approach as opposed to conventional medical ultrasonic transducers with backing layers. The electronics module comprises a microcontroller unit (MCU), along with pulser and receiver circuitry for excitation of the transducers, reception of echo signals and communication with a mobile device. From the electronics design point of view, the UBVM device bypasses power-hungry data acquisition modules by utilizing a 1-bit analog-to-digital conversion scheme as a simple yet efficient time-of-flight measurement strategy (Supplementary Note 3). Such a strategy not only decreases the power consumption of the device but also eliminates additional ICs and peripheral components for the miniaturization of the system. Finally, the custom mobile app communicates with the UBVM hardware every 25 ms for real-time bladder volume monitoring. In addition, the app is capable of communicating with Amazon Web Services (AWS) for potential integration into existing cloud-based patient monitoring services.

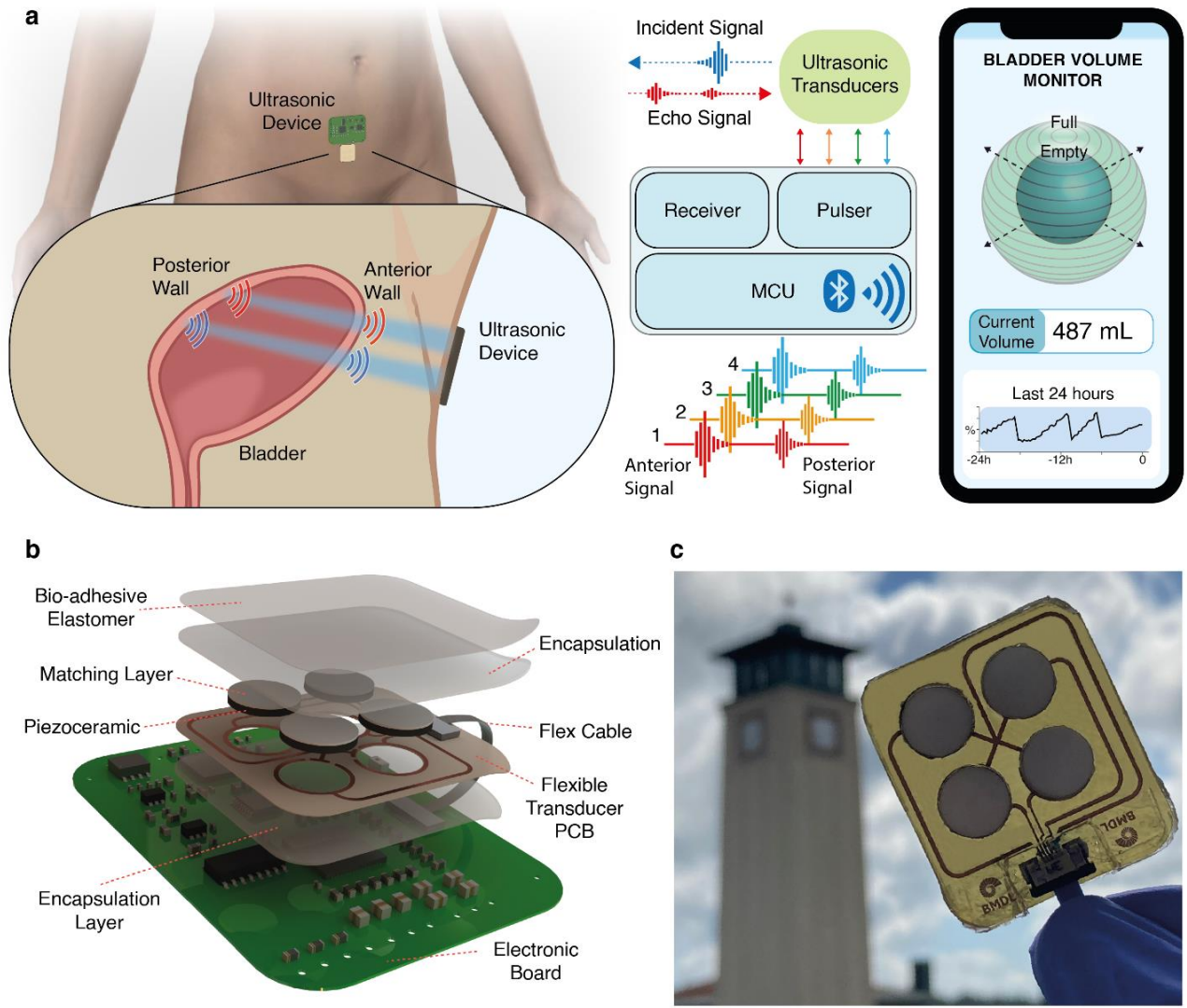

***Fig.1. Design and working principle of the integrated ultrasonic device. a**, Schematic of the UBVM device placed on the lower abdomen with key functionalities. The device enables continuous bladder volume monitoring by continuously measuring the anterior-posterior (A-P) wall distance of the bladder from multiple sites and processing them to estimate the volume. Within the UBVM hardware, the MCU constantly communicates with the pulser and receiver modules to excite the ultrasonic transducers and record the received echoes from anterior and posterior walls of the bladder. Received echo timestamps are transferred to a custom-made app for processing and visualization. **b**, Exploded view of the UBVM device, including the printed circuit board (PCB) and layers of the ultrasonic transducers. **c**, Photograph of the ultrasonic transducers.*

**Characterization of ultrasonic transducers.** Conventional ultrasonic transducers used in medical imaging contain backing layers made up of absorptive materials to damp out the ringing effect, which sacrifices the sensitivity as most of the energy is absorbed by the backing layer (Supplementary Fig. 1, 2) [49]. To enable a simple, small footprint and efficient wearable platform, we have opted for an air-backed design with a wrap-around electrode configuration where the former improves the sensitivity of the transducer dramatically and allow measurements with low driving voltages and the latter simplifies the fabrication procedure by enabling electrode access from the same plane (Supplementary Note 4). Fig. 2a shows the cross-sectional view of the transducer with structural layers and the air cavity. By using a commercially available copper/polyimide (Cu, 9 μm / PI, 25 μm) flexible laminate, a 40 μm thick polyethylene terephthalate (PET) encapsulation layer in the back, a thin (100 μm) polydimethylsiloxane (PDMS) encapsulation layer in the front and a silicone-based adhesive, the UBVM transducers provide conformable coupling to the skin and enable gel-free measurements (Supplementary Fig. 3). In order to compensate for the narrower bandwidth associated with the air-backed design and improve the transducer performance, we have also included a matching layer in front. With a delicate laser etching step, we were able to optimize the matching layer thickness in micron-level precision for optimum transducer performance. (Fig. 2b, Supplementary Note 5, 6)

Considering the frequency constraints of the MCU-generated pulses, the deep position of the posterior wall of the bladder and the frequency-dependent attenuation of ultrasound waves in tissue[49], 2 MHz was selected as the operating frequency of the transducer. Fig. 2c shows the pulse-echo response of the transducer and its derived frequency spectrum, which shows the resonance frequency of the transducer 2 MHz and reveals a -3 dB bandwidth of 29.1%. Both the electrical impedance and phase angle measurements (Fig. 2d) and the underwater excitation frequency sweep of the transducer (Fig. 2e) also show the 2 MHz resonance frequency of the device, in agreement with each other and the pulse-echo response. To highlight the influence of air-backing, the pulse-echo response of the UBVM device transducer was compared to a commercially available immersion transducer with an operating frequency of 2.25 MHz (Olympus V232-SU) in Fig. 2f. Even though the commercial transducer had a larger active area, the UBVM device transducer demonstrated about 4-fold increase in SNR in comparison to the commercial transducer. The measured 2D pressure field of the transducer in the longitudinal direction is shown in Fig. 2g, showing the directivity of the transducer as a result of the chosen aspect ratio of the PZT disc. This is further supported by the transverse pressure field scan of the transducer, where the -6 dB contour lines are smaller than the radius of the PZT disc (Fig. 2h and Supplementary Fig. 4).

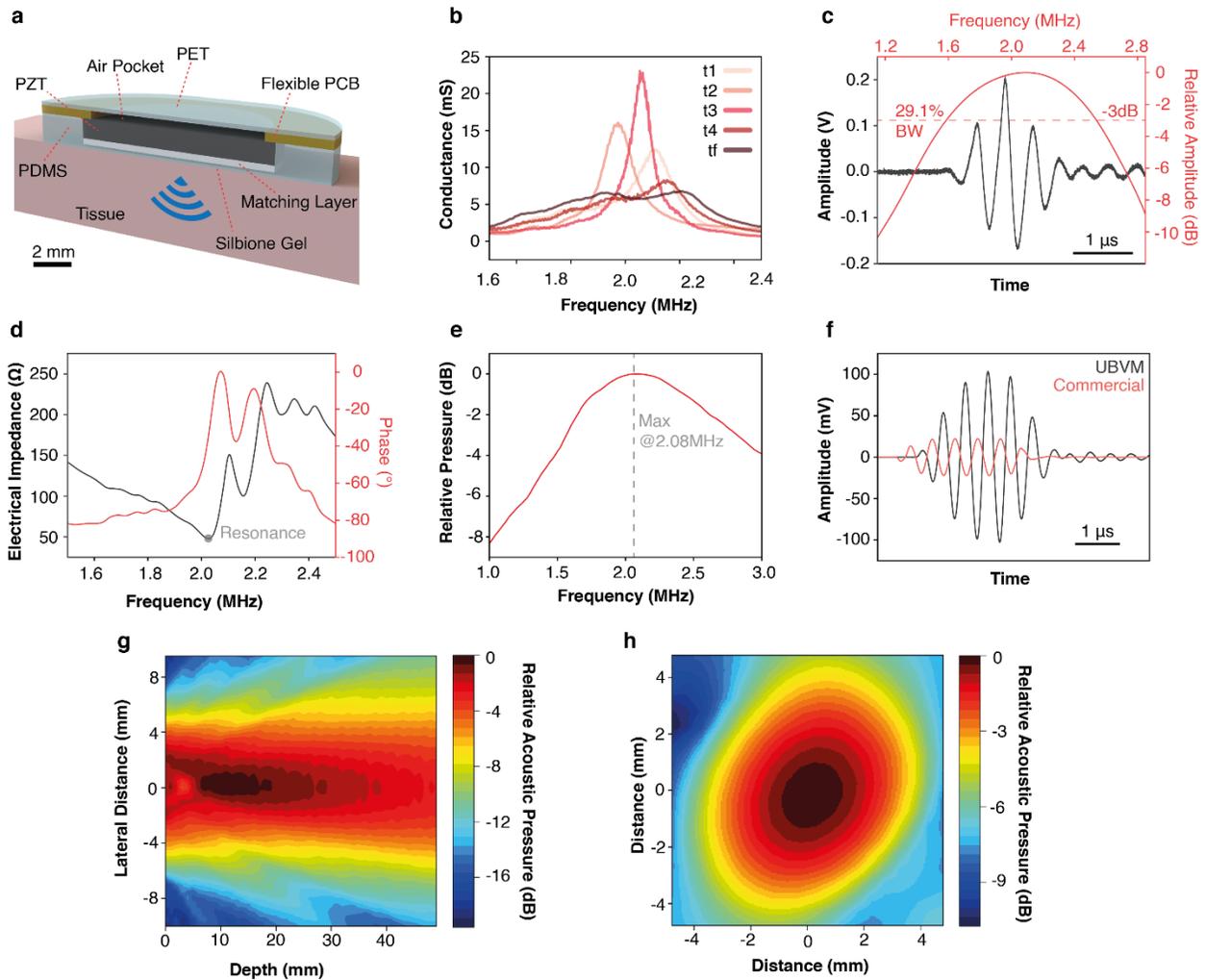

***Fig.2. Characterization of the ultrasonic device. a**, Cross-sectional view of the ultrasonic transducers including each layer and the air-backing. Air-backed design eliminates energy loss due to absorptive backing layers and allows ultrasonic sensing with lower voltages. **b**, Optimization of the matching layer thickness by conductance vs frequency plots. The matching layer was etched until the thickness tf, corresponding to the maximum flat conductance of the transducer. **c**, Pulse-echo response and the derived frequency spectra of a transducer, highlighting the resonance frequency of the transducer. **d**, Impedance and phase angle spectra of a transducer, showing the minimum impedance value at the expected resonance frequency. **e**, Frequency sweep of the transducer captured by a hydrophone. Dashed line represents -3dB bandwidth of the transducer. **f**, Echo signal level comparison of the UBVM air-backed transducer to a commercial transducer with backing layer, showing the enhanced sensitivity of our design. **g**, Mapped pressure field of a UBVM transducer, highlighting the directivity of transducer. **h**, Cross-sectional field map of the insonation region, parallel to the surface of the transducer.*

**Electronics design and in-vitro characterization of the system.** Fig. 3a shows a real photo of the transducers with the fabricated PCB. The system uses six integrated circuits (ICs) to excite the transducers, acquire the received signals and transmit the recorded timestamps (Fig. 3b). As illustrated in Fig. 3c, the microcontroller unit generates the trigger pulses and multiplexer channel selection codes to address an individual transducer. The multiplexer then transfers the trigger pulses to the chosen channel of the pulser. The pulser generates high voltage pulses for transducer excitation and utilizes its embedded transmit & receive switch to forward the received echo to the receive circuitry. Another multiplexer, controlled by the same selection codes from the MCU, then directs the echo signal, passing through a set of filters and amplifiers, to the comparator. The output of the comparator is sampled by the MCU to acquire the timestamps of the echo signals, which are sent to a mobile phone with a custom app for volume estimation (Supplementary Note 3).

To evaluate the performance of the electronics module, a pulse-echo experiment was conducted in a water tank with commercial transducers. By adapting the clock of our MCU, we were able to excite several commercial ultrasonic transducers at their respective frequencies, and wirelessly transfer the echo timestamps to our custom app in real-time. A steel reflector was positioned at different distances from the transducers to simulate the anterior and posterior wall distances. As depicted in Fig. 3d, the UBVM electronics successfully conducted wireless pulse-echo measurements across the medically relevant frequency range. Once the performance of the UBVM electronics module was validated, the full UBVM device was characterized in vitro in a water tank where the bladder was mimicked by the spherical-shaped round-bottom flasks (Fig 3e). Measurements were conducted in deionized water, which has a similar acoustic impedance to that of tissue[50]. 250 mL and 500 mL round-bottom flasks were used as the test objects. As seen in Fig. 3f, the UBVM device estimated the flask volumes as 274 and 547 mL, respectively, yielding an error of less than 10%.

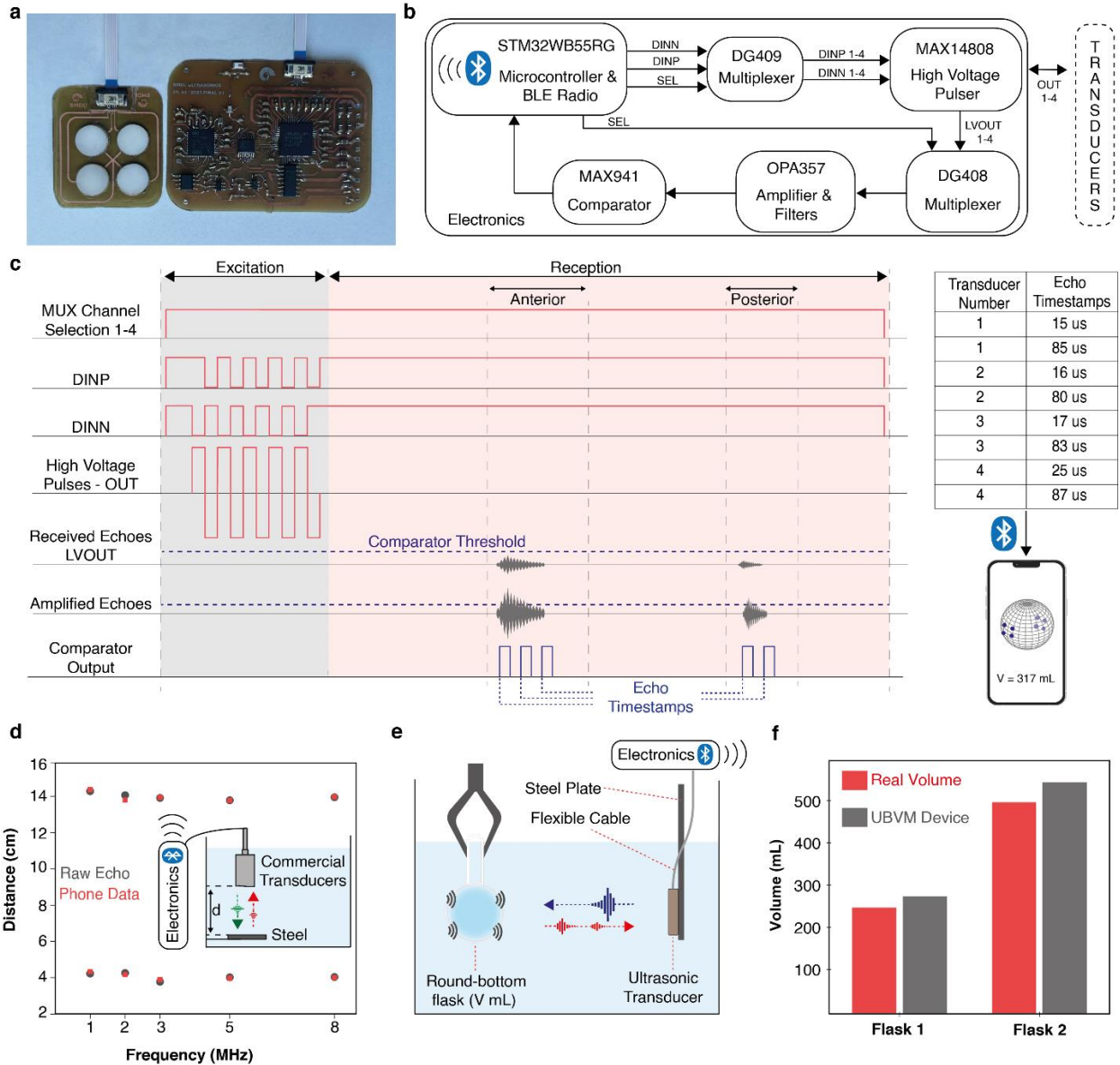

***Fig.3. Electronics design and in-vitro demonstration of the UBVM device. a***, Photograph of the PCBs with soldered components. ***b***, Functional block diagram of signal transmission lines of the UBVM device electronics. ***c***, Timing diagram of the UBVM electronics. A transducer channel is selected by the MCU and the MUX. 2 sets of unipolar pulses with 180° phase difference allow the pulser to generate bipolar high voltage pulses. Received echoes are passed to the receiver circuitry via the internal switch of the pulser. Echoes are amplified, passed through the comparator for echo timestamp detection and transferred to the mobile phone for processing and visualization. ***d***, Schematic drawing of the pulse-echo setup to demonstrate the wireless pulse-echo capability at various frequencies. Echo timestamps displayed on the mobile phone exactly correspond to the raw echo signals on the oscilloscope. ***e***, Schematic drawing of the in-vitro setup of the whole system. The test set-up mimics the bladder with round-bottom flasks in deionized water, where the transducers were connected to the electronics with a 1-meter-long flex cable. ***f***, Comparison of the measured flask volumes with actual volumes, with a less than 10% error range on spherical shaped objects.

**In-vivo validation of continuous bladder volume measurement capabilities.** In-vivo measurements were conducted for further demonstration of the UBVM device's continuous bladder volume monitoring capability. Fig. 4a shows a photograph of the transducers on the volunteer's lower abdomen. The custom mobile app not only processes the measurement data and visualizes them, but also guides the user for optimal placement of the transducers. In an ideal measurement, the transducers should receive eight echo signals in total, four from the anterior and another four from the posterior wall of the bladder. The app analyzes the received signals and alerts the user if the number of echoes received is below five, which is the minimum required for accurate spherical fitting, prompting the user to reposition the transducers.

In order to observe a full micturition cycle, measurements were initiated with an empty bladder and the transducers are placed with guidance from the mobile app. Volume estimations were recorded every 30 minutes by opening the custom app for automatic calculation of the bladder volume. Fig. 4d shows a full micturition cycle of the volunteer captured by the UBVM device and the A-P distance data from one of the transducers. The increasing bladder volume trend can easily be seen. The UBVM transducers measure the A-P wall distances of the bladder, where multiple distance measurements are used to estimate the volume.

To validate the accuracy of the volume estimates, ultrasound images of the bladder were used to calculate the bladder volumes of five healthy (3 females and 2 males) volunteers (Fig .4b,c and Supplementary Fig. 5). The ultrasound images were captured by an experienced urologist using a commercial ultrasound imaging system. After the ultrasonographic bladder volume assessments of the participants, the values were compared to the UBVM device estimates. As shown in Fig. 4e, the UBVM device accurately estimated the bladder volumes of the participants with volumes ranging from as low as 84 mL to as high as 800 mL, demonstrating its performance in various bladder shapes and sizes.

Raw echo signals and the associated comparator output are shown in Fig. 4f and Fig. 4g. Thanks to the air-backed design with the matching layer, echoes from the anterior and posterior walls of the bladder are clearly captured, even with an excitation voltage as low as 30 V.

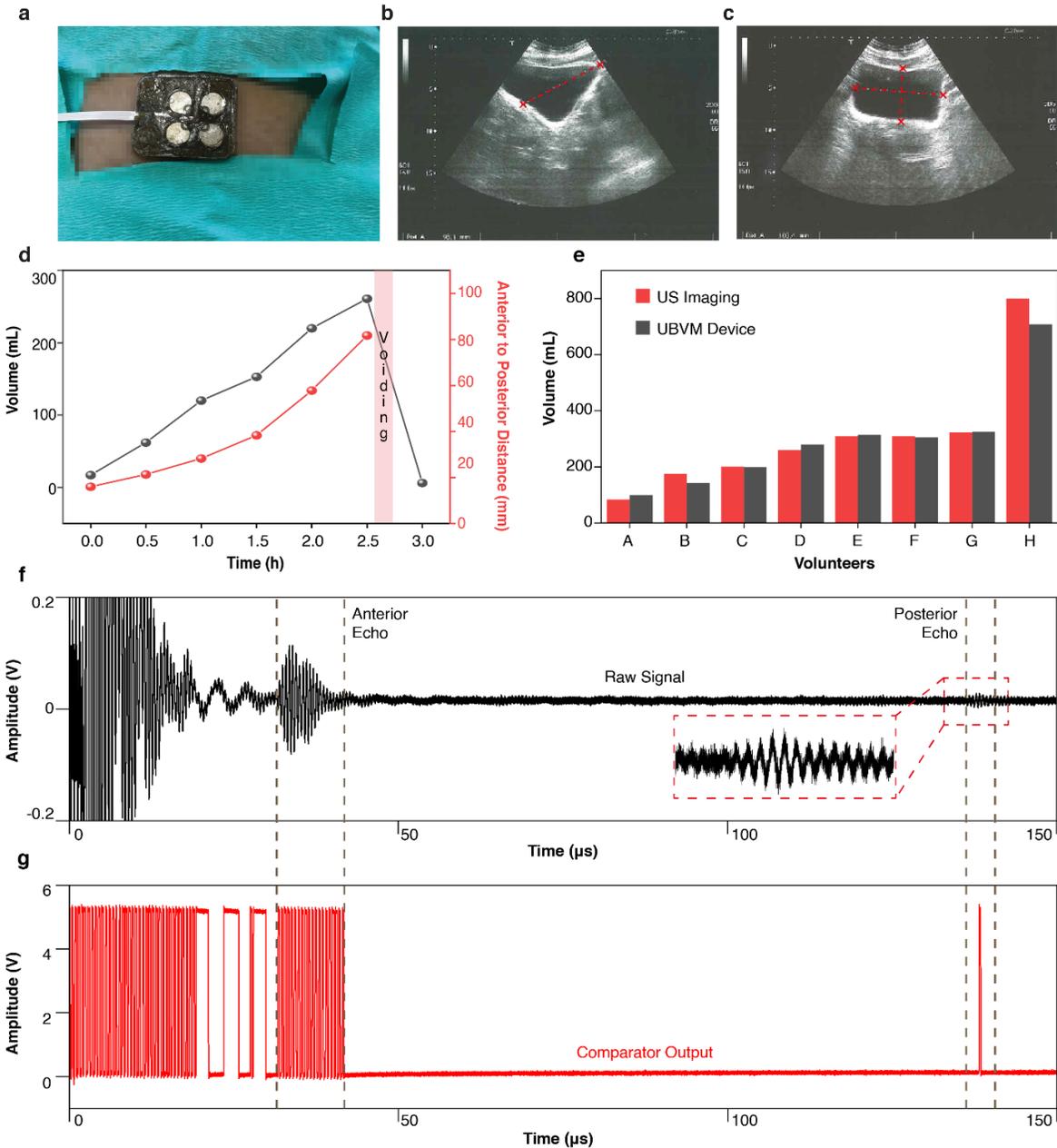

*Fig.4. In-vivo validation of continuous bladder volume measurement capabilities. a, Image of the transducers on lower abdomen of a volunteer. b, Ultrasound image of a volunteer's bladder with transverse probe orientation. c, Ultrasound image of a volunteer's bladder with longitudinal probe orientation. d, Continuous monitoring during a full micturition cycle (red) and the A-P distance measurements of a transducer. Multiple A-P distance measurements from multiple transducers allow spherical volume estimation. e, Comparison of the UBVM device measurements with the volumes from conventional ultrasound imaging. A, B, D and G represent measurements from the same volunteer at different time points, while C, E, F and H correspond to measurements from different volunteers. f, Raw echo signals recorded on an oscilloscope during an in-vivo measurement. Both the anterior and the posterior echo can be seen. g, Signal as captured by the MCU after receiver circuitry. These timestamps are used for bladder volume estimations.*

**Discussion**

We have described a flexible and integrated ultrasonic system enabling continuous measurement of the bladder volume. The UBVM device combines a flexible transducer module with an air-backed design and a simple fabrication methodology that enables low driving voltages, an electronics system enabling low-power and miniaturized A-scan measurements with a custom mobile app and data processing strategies for accurate bladder volume measurements. While our focus has been on the bladder volume monitoring, extended versions of the system with minor hardware and firmware modifications can open up new dimensions in wearable ultrasound imaging as shown in recent studies[48].

Future work will focus on increasing the number of transducers while reducing the overall footprint. This can be achieved by increasing the frequency of the transducers and shrinking their size in all three dimensions and expanding the number of channels of the multiplexers used in the current version of the UBVM electronics. In addition, since both the transducers and the electrical components are placed on printed circuit boards, an electronics and transducers co-design at the PCB level could create room for further miniaturization. Once these challenges are addressed, we anticipate the UBVM device to be an invaluable tool in addressing various use-cases, such as lower urinary tract symptoms and post-operative urinary retention.

**Methods**

**Preparation of the PZT transducers.** Bulk PZT discs were purchased from American Piezo (APC - 850). To enable access to top and bottom electrodes from the same plane, discs were modified to a custom wrap-around electrode configuration. A commercial picosecond-pulsed laser ablation system (Supplementary Fig. 6; 500 kHz pulse repetition frequency with a divider of 3, 1.5 W power and 1500 mm s−1 laser cutting speed, R4, LPKF Laser & Electronics) was utilized to isolate a small portion of the silver electrode on one face of the PZT disc. Then, the isolated portion of the silver electrode on that face was electrically connected to the electrode on the opposite face of the disc using a low temperature silver paste (PE827, DuPont).

**Fabrication of the flexible ultrasonic transducers.** 25 µm thick flexible laminate with 9 µm copper layer (AC0925, DuPont) was patterned with LPKF U4 to create the circuit pattern of the transducers and cut the laminate for air-backing. Two-component epoxy resin (EPOTEK 301-2FL, 1:0.35) was mixed at 3500 rpm for 20 minutes using a speed mixer (Hauschild mixer, DAC 150). $Al_2O_3$ nanoparticles (290nm, Nanografi) were added to the mixture for desired acoustic impedance and mixed at 3500 rpm for another 10 minutes using the same speed mixer. The epoxy-nanoparticle mixture was poured inside a disc shaped PDMS (Dow Corning, 10:1) mold. Cured matching layer was manually bonded to PZT discs using a fast curing two-component epoxy (Loctite EA3 422, 1:1). PZT discs with matching layers were manually placed and reflowed with low-temperature solder paste (TS391LT, Chip Quik) (Supplementary Fig. 7 and 8).

**Matching layer thickness optimization.** After identifying the matching layer composition with desired density and speed of sound (Supplementary Fig. 9), LPKF R4 was utilized (300 kHz pulse repetition frequency with a divider of 2, 7.3 W power and 1400 mm s−1 laser cutting speed) to thin the matching layer into desired thickness. Laser parameters were optimized to remove a micrometer of matching layer in each pass. After each material removal step, conductance measurements were performed (MFIA Impedance Analyzer, Rohde & Schwarz) in deionized water to find the optimum matching layer thickness. Laser etching followed by admittance measurement procedure was repeated until the broadest conductance measurement was obtained.

**Packaging.** To preserve the air-backed design of the transducers, laminate under the disc-shaped PZTs were cut. In an analogous fashion, a double-sided tape (Tesa 64621) was machined. PZT-soldered laminate was adhered to the bottom encapsulation layer (40 µm thick PET sheet) using the double-sided tape. For the front side encapsulation, a mold was 3D printed (BMF microArch S230) and parylene-c coated (Plasma Parylene Systems). The laminate with transducers were placed inside the mold. Silicone encapsulant (PDMS, 10:1) was poured and cured at 60°C for 9 hours. To enable gel-free measurements, Silbione 4717 gel (1:1) was screen-printed on the front side of the transducers (Supplementary Fig. 10 and 11).

**Characterization of the transducers.** All electrical and acoustics characterizations were conducted in deionized water. An impedance analyzer (MFIA, Zurich Instruments) was employed to obtain frequency dependent electrical parameters (electrical impedance, phase, conductance) of the transducer. For the pulse-echo response, an arbitrary waveform generator (33521B Waveform Generator, Keysight) was used to excite the transducer-under-test. The echo signal from a stainless steel reflector was received by an osciloscope (Picoscope 6000 series) through the use of a diplexer (RDX-6, RITEC Inc.) (Supplementary Fig. 12).

Underwater characterizations were conducted inside an UMS Research system covered with acoustic absorbers (AptFlex F28, Precision Acoustics). The pressure field, sound pressure output and bandwidth were evaluated with a 0.5 mm needle hydrophone (Precision Acoustics) connected to an oscilloscope (DSOX3024G, Keysight Technologies).

**Electronics system design and fabrication.** Modeling and simulation of the UBVM hardware were performed in LTSpice (Supplementary Fig. Y8). Following the simulations, all components were purchased from Digikey Electronics. Double-sided printed circuit board fabrication started with drilling via holes on copper-clad laminates as per design (LPKF Protomat E44). Drilling step was followed up by pulsed direct current electroplating of copper to connect the top and bottom layers through drilled holes (LPKF Contac S4). Both sides were then patterned using LPKF U4. As shown in Supplementary Fig. 20, main components include a Bluetooth Low Energy integrated microcontroller (STM32WB55RG, STMicroelectronics), a single and a dual channel multiplexer (DG408 and DG409, Maxim Integrated), a high voltage pulser (MAX14808, Maxim Integrated), an operational amplifier (OPA357, Texas Instruments) and a comparator (MAX941, Maxim Integrated) (Supplementary Fig. 13-20).

**Transmit Phase.** During the transmit phase, the microcontroller generates two sets of five 2 MHz pulses amplitude of 3.3V every 2,5 seconds (Supplementary Fig. 21). The two sets have a 180° phase difference and are required by the pulser IC for bipolar pulse generation. Through the use of the multiplexer, four individual input channels of the high-voltage pulser are switched at

10-second intervals. Ultrasonic transducers are then excited by bipolar pulses as per the input excitation signals and high voltage DC supply level of the pulser (Supplementary Fig. 22).

**Receive Phase.** The MAX14808 high-voltage pulser includes a transmit/receive switch, which forwards the received signal from bladder walls to the receive circuitry. The op-amp provides a gain of 10 with 1k and 10k resistors, followed by an RC low-pass filter with a -3 dB cut-off frequency of 5.88 MHz to eliminate the higher order harmonics in the received signal caused by the gain bandwidth product of the amplifiers of the amplifiers. The amplified signal is then fed to the comparator, which is used as a threshold detector. If the received and amplified signal surpasses the comparator threshold (e.g. when an echo is received) the comparator output goes high (5 V). Otherwise, the comparator output is low (0 V) (Supplementary Fig. 23). To block false firings at the output, comparator threshold is dynamically varied through the use of a simple resistive network composed of two 20k resistors that is series between the supply and the ground, the divided voltage then fed back to the comparator's positive pin over a 60k resistor. The ratio of these resistors determines the amount of the hysteresis.

**Data Acquisition and Bluetooth Operation.** Proposed one-bit analog to digital conversion allows the use of the input capture mode of the microcontroller. More specifically, a timer channel of the microcontroller is configured to capture the value of a counter at the moment the input signal transitions from one state to another. By connecting the timer channel to the comparator output, time-stamps of the echo signals are detected and stored by the microcontroller through the use of direct memory access (DMA). This storage process provides fast capturing of the incoming data without interrupting the work of the microcontroller. Subsequently, this data is transmitted via Bluetooth Low Energy consisting of a Generic Attribute Profile (GATT) server that includes custom services and characteristics. In the context of capturing echo timestamps, one service that contains a data capturing parameter as a characteristic and transfers the data over BLE was used. This characteristic can be accessed by other BLE devices, allowing them to read, write, or be notified. The data send over BLE is then passed through a custom application on a mobile phone for further processing in Amazon Web Services (AWS). The phone between the wireless microcontroller and AWS behaves as a gateway for conveying the data using the Internet. The data processing stages involved in AWS allow for the transmission of bladder volume information to the phone, passing through the following services: AWS IoT, AWS Kinesis Data Firehose, AWS S3 Bucket, AWS Lambda, AWS AppSync, AWS Dynamo Table, and finally, the user application. GraphQL language is employed in this pipeline to facilitate interaction between users and the services involved. Users can utilize GraphQL's capabilities to perform operations such as reading and writing data, as well as connecting their smartphones to the pipeline through an Application Programming Interface (API).

**Cloud Data Processing.** The timestamps captured by the microcontroller are processed in AWS Lambda service using Python. Each timestamp arriving at AWS Lambda triggers the Python code. Therefore, the algorithm continuously registers the timestamps into the data array and checks if the monitoring is complete based on the transducer number information attached to the timestamps. Once the transducer number changes from 4 to 1, the algorithm understands that the monitoring is complete and starts processing the collected data to approximate the bladder volume. First, the timestamp array is masked to determine the number of data points that can be used in the volume calculation. Since a sphere can be defined by a minimum of 4 points, if there is insufficient number of data points in the array, the algorithm sends an error message to the mobile app. Otherwise, the processing continues with the averaging step. Timestamps for anterior and posterior signals of each transducer are averaged and registered to a new array. Then timestamps in the averaged data array are converted to coordinates assuming a speed of sound of 1480 m/s[50]. Finally, the coordinate array is fed into the sphere fitting function, which finds the volume of the best-fit sphere using least squares method that utilizes the Broyden-Fletcher-Goldfarb-Shanno (BFGS) minimization algorithm.

**In vitro characterization.** The stainless-steel reflector and round bottom flasks in different volumes were used to evaluate the performance of the developed system in vitro. To test the capabilities of our electronics and firmware, the pulse-echo experiments were conducted at different frequencies. The stainless-steel reflector was attached to a motorized stage of the UMS Research system. Commercial ultrasonics transducers with various operating frequencies (1-2-3-5-8 MHz) were submerged into UMS Research system and were connected to the developed electronics. The transducers were excited, and the received echoes were captured by our electronics. Captured echo timestamps were then transferred to a mobile phone.
To mimic our in-vivo application, two round bottom flasks were submerged into UMS Research system, along with our transducers. A flex cable was used to connect the transducers to the PCB outside the water tank. Received timestamps were transferred to a mobile phone and converted into coordinate points and fitted to a spherical shape. After the calculations, the volume data of the flask-under-test was displayed.

**In vivo measurements.** Continuous bladder volume measurement was carried out on five healthy volunteers with an approved protocol by Koc University Ethics Committee (2023.006.IRB2.004). The UBVM device was placed on the lower abdomen just above the pubic bone for the measurements. For comparison with clinical ultrasound imaging, the volunteers were examined in supine position with Toshiba Nemio 10 scanner equipped with a 3.5 MHz convex array transducer by an experienced urologist. Bladder images representing the maximum longitudinal, transverse and anterior-posterior diameters were recorded. The diameter dimension were determined by manually placing the calipers on the recorded images. Based on the measured dimensions, spherical volume formula with a correction coefficient of 0.52 was used to calculate the bladder volume[51].

**Code availability**

The firmware of the electronic hardware, written in C, the data processing code, written in Python, and the source code for app, written in Javascript, are available from the authors for research purposes on request.

**Acknowledgements**

We thank Gokberk Toymus for discussions on lower urinary tract dysfunctions, Seckin Akinci and Fariborz Mirlou for the help with PCB design and fabrication, Mohammad Meraj Ghanbari for the help with the pulser, Prof. Dr. Hayrettin Koymen for discussions on ultrasonic transducers, Ecem Ezgi Ozkahraman for the help with in-vivo measurements and the feedback on the manuscript.

A.T.T is supported by Scientific and Technological Research Council of Turkey (TUBITAK) through the 2232 (grant no. 118C295) programme. U.C.Y. is supported by TUBITAK through the 2210/A programme. E.B. is supported by TUBITAK through the 2210/C programme. L.B. acknowledges TUBITAK 2232 (grant no. 118C295) and the European Research Council (grant no. 101043119). We acknowledge Koç University Surface Science and Technology Center (KUYTAM) and Koç University Nanofabrication and Nanocharacterization Center (n2STAR) for access to the infrastructure.

**References**

1. D'Ancona, C. *et al.* The International Continence Society (ICS) report on the terminology for adult male lower urinary tract and pelvic floor symptoms and dysfunction. *Neurourol Urodyn* **38**, 433–477 (2019).
2. Griebling, T. L. WORLDWIDE PREVALENCE ESTIMATES OF LOWER URINARY TRACT SYMPTOMS, OVERACTIVE BLADDER, URINARY INCONTINENCE, AND BLADDER OUTLET OBSTRUCTION. *BJU Int* **108**, 1138–1139 (2011).
3. Awedew, A. F. *et al.* The global, regional, and national burden of benign prostatic hyperplasia in 204 countries and territories from 2000 to 2019: a systematic analysis for the Global Burden of Disease Study 2019. *Lancet Healthy Longev* **3**, e754–e776 (2022).
4. Phé, V., Chartier–Kastler, E. & Panicker, J. N. Management of neurogenic bladder in patients with multiple sclerosis. *Nat Rev Urol* **13**, 275–288 (2016).
5. Panicker, J. N., Fowler, C. J. & Kessler, T. M. Lower urinary tract dysfunction in the neurological patient: clinical assessment and management. *Lancet Neurol* **14**, 720–732 (2015).
6. Launer, B. M., McVary, K. T., Ricke, W. A. & Lloyd, G. L. The rising worldwide impact of benign prostatic hyperplasia. *BJU Int* **127**, 722–728 (2021).
7. Nieuwhof-Leppink, A. J., Schroeder, R. P. J., van de Putte, E. M., de Jong, T. P. V. M. & Schappin, R. Daytime urinary incontinence in children and adolescents. *Lancet Child Adolesc Health* **3**, 492–501 (2019).
8. Gammie, A. *et al.* What developments are needed to achieve less-invasive urodynamics? ICI-RS 2019. *Neurourol Urodyn* **39**, S36–S42 (2020).
9. Smith, A., Bevan, D., Douglas, H. R. & James, D. Management of urinary incontinence in women: summary of updated NICE guidance. *BMJ* **347**, (2013).
10. Bright, E., Cotterill, N., Drake, M. & Abrams, P. Developing and Validating the International Consultation on Incontinence Questionnaire Bladder Diary. *Eur Urol* **66**, 294–300 (2014).
11. Mickle, A. D. *et al.* A wireless closed-loop system for optogenetic peripheral neuromodulation. *Nature 2019 565:7739* **565**, 361–365 (2019).
12. Hannah, S., Brige, P., Ravichandran, A. & Ramuz, M. Conformable, Stretchable Sensor to Record Bladder Wall Stretch. *ACS Omega* **4**, 1907–1915 (2019).
13. Stauffer, F. *et al.* Soft Electronic Strain Sensor with Chipless Wireless Readout: Toward Real-Time Monitoring of Bladder Volume. *Adv Mater Technol* **3**, 1800031 (2018).
14. Mandal, S., Dehghanzadeh, P., Zamani, H. & Shaik, S. A Wireless Implantable Microsystem for Real-Time Bladder Volume Monitoring. (2022).
15. Gerber, M. T. Impedance-based bladder sensing. Preprint at (2015).

16. Thüroff, J. W. *et al.* EAU Guidelines on Urinary Incontinence. *Eur Urol* **59**, 387–400 (2011).
17. Abelson, B. *et al.* Ambulatory urodynamic monitoring: state of the art and future directions. *Nature Reviews Urology 2019 16:5* **16**, 291–301 (2019).
18. SENS-U KIDS Bladder Sensor is our first proof of wearable ultrasound care. https://novioscan.com/.
19. DFree - US site wearable bladder scanner for incontinence. https://www.dfreeus.biz/.
20. Lilium Otsuka Co., Ltd. https://www.lilium.otsuka/en/.
21. van Leuteren, P. G., Nieuwhof-Leppink, A. J. & Dik, P. SENS-U: clinical evaluation of a full-bladder notification – a pilot study. *J Pediatr Urol* **15**, 381.e1-381.e5 (2019).
22. Hofstetter, S. *et al.* Dfree ultrasonic sensor in supporting quality of life and patient satisfaction with bladder dysfunction. *International Journal of Urological Nursing* **17**, 62–69 (2023).
23. Kamei, J., Watanabe, D., Homma, Y., Kume, H. & Igawa, Y. Feasibility of approximate measurement of bladder volume in male patients using the Lilium α-200 portable ultrasound bladder scanner. *LUTS: Lower Urinary Tract Symptoms* **11**, 169–173 (2019).
24. Niestoruk, L. *et al.* A concept for wearable long-term urinary bladder monitoring with ultrasound. Feasibility study. *EDERC 2012 - Proceedings of the 5th European DSP in Education and Research Conference* 134–138 (2012) doi:10.1109/EDERC.2012.6532241.
25. Semproni, F., Iacovacci, V. & Menciassi, A. Bladder Monitoring Systems: State of the Art and Future Perspectives. *IEEE Access* **10**, 125626–125651 (2022).
26. Kiczek, B. *et al.* A wearable ultrasonic bladder monitoring device. *Proceedings of the Annual International Conference on Mobile Computing and Networking, MOBICOM* 886–888 (2022) doi:10.1145/3495243.3558272.
27. Jo, H. G. *et al.* Forward-looking ultrasound wearable scanner system for estimation of urinary bladder volume. *Sensors* **21**, 5445 (2021).
28. Fournelle, M. *et al.* Portable Ultrasound Research System for Use in Automated Bladder Monitoring with Machine-Learning-Based Segmentation. *Sensors 2021, Vol. 21, Page 6481* **21**, 6481 (2021).
29. Gao, Q. *et al.* Design and Development of a Bladder Volume Determination Device Based on A-mode Ultrasound. *2021 IEEE International Conference on Mechatronics and Automation, ICMA 2021* 170–175 (2021) doi:10.1109/ICMA52036.2021.9512651.
30. Kuru, K. *et al.* Intelligent autonomous treatment of bedwetting using non-invasive wearable advanced mechatronics systems and MEMS sensors: Intelligent autonomous bladder monitoring to treat NE. *Med Biol Eng Comput* **58**, 943–965 (2020).
31. Gaubert, V., Gidik, H. & Koncar, V. Smart underwear, incorporating textrodes, to estimate the bladder volume: proof of concept on a test bench. *Smart Mater Struct* **29**, 085028 (2020).
32. Shin, S. C. *et al.* Continuous bladder volume monitoring system for wearable applications. *Proceedings of the Annual International Conference of the IEEE Engineering in Medicine and Biology Society, EMBS* 4435–4438 (2017) doi:10.1109/EMBC.2017.8037840.
33. Schlebusch, T., Nienke, S., Leonhardt, S. & Walter, M. Bladder volume estimation from electrical impedance tomography. *Physiol Meas* **35**, 1813 (2014).
34. Reichmuth, M., Schurle, S. & Magno, M. A Non-invasive Wearable Bioimpedance System to Wirelessly Monitor Bladder Filling. *Proceedings of the 2020 Design,*

*Automation and Test in Europe Conference and Exhibition, DATE 2020* 338–341 (2020) doi:10.23919/DATE48585.2020.9116378.
35. Fong, D., Alcantar, A. V., Gupta, P., Kurzrock, E. & Ghiasi, S. Non-invasive bladder volume sensing for neurogenic bladder dysfunction management. *2018 IEEE 15th International Conference on Wearable and Implantable Body Sensor Networks, BSN 2018* **2018-January**, 82–85 (2018).
36. Molavi, B., Shadgan, B., Macnab, A. J. & Dumont, G. A. Noninvasive optical monitoring of bladder filling to capacity using a wireless near infrared spectroscopy device. *IEEE Trans Biomed Circuits Syst* **8**, 325–333 (2014).
37. FechnerPascal, KönigFabian, KratschWolfgang, LocklJannik & RöglingerMaximilian. Near-Infrared Spectroscopy for Bladder Monitoring: A Machine Learning Approach. *ACM Trans Manag Inf Syst* **14**, 1–23 (2023).
38. Koven, A. & Herschorn, S. NIRS: Past, Present, and Future in Functional Urology. *Curr Bladder Dysfunct Rep* **17**, 241–249 (2022).
39. Leonhäuser, D. *et al.* Evaluation of electrical impedance tomography for determination of urinary bladder volume: Comparison with standard ultrasound methods in healthy volunteers. *Biomed Eng Online* **17**, 1–13 (2018).
40. Hu, H. *et al.* Stretchable ultrasonic transducer arrays for three-dimensional imaging on complex surfaces. *Sci Adv* **4**, (2018).
41. Wang, C. *et al.* Monitoring of the central blood pressure waveform via a conformal ultrasonic device. *Nature Biomedical Engineering 2018 2:9* **2**, 687–695 (2018).
42. Wang, F. *et al.* Flexible Doppler ultrasound device for the monitoring of blood flow velocity. *Sci Adv* **7**, 9283–9310 (2021).
43. Wang, C. *et al.* Continuous monitoring of deep-tissue haemodynamics with stretchable ultrasonic phased arrays. *Nature Biomedical Engineering 2021 5:7* **5**, 749–758 (2021).
44. Wang, C. *et al.* Bioadhesive ultrasound for long-term continuous imaging of diverse organs. *Science (1979)* **377**, 517–523 (2022).
45. Hu, H. *et al.* A wearable cardiac ultrasound imager. *Nature 2023 613:7945* **613**, 667–675 (2023).
46. Hu, H. *et al.* Stretchable ultrasonic arrays for the three-dimensional mapping of the modulus of deep tissue. *Nature Biomedical Engineering 2023* 1–14 (2023) doi:10.1038/s41551-023-01038-w.
47. Yu, C.-C. *et al.* A Conformable Ultrasound Patch for Cavitation-Enhanced Transdermal Cosmeceutical Delivery. *Advanced Materials* 2300066 (2023) doi:10.1002/ADMA.202300066.
48. Lin, M. *et al.* A fully integrated wearable ultrasound system to monitor deep tissues in moving subjects. *Nature Biotechnology 2023* **7**, 1–10 (2023).
49. Shung, K. K. & Zipparo, M. J. Ultrasonic transducers and arrays. *IEEE Engineering in Medicine and Biology Magazine* **15**, 20–30 (1996).
50. Kino, G. S. Acoustic Waves: Devices, Imaging, and Analog Signal Processing. *Englewood Cliffs* **100**, 103 (1987).
51. Bih, L. I., Ho, C. C., Tsai, S. J., Lai, Y. C. & Chow, W. Bladder shape impact on the accuracy of ultrasonic estimation of bladder volume. *Arch Phys Med Rehabil* **79**, 1553–1556 (1998).